\documentclass[conference]{IEEEtran}
\IEEEoverridecommandlockouts
\usepackage{cite}
\usepackage{amsmath,amssymb,amsfonts}
\usepackage{algorithmic}
\usepackage{graphicx}
\usepackage{textcomp}
\usepackage{multirow}
\usepackage[table,xcdraw]{xcolor}
\def\BibTeX{{\rm B\kern-.05em{\sc i\kern-.025em b}\kern-.08em
    T\kern-.1667em\lower.7ex\hbox{E}\kern-.125emX}}
\begin{document}

\title{TSI-Net: A Timing Sequence Image Segmentation Network for Intracranial Artery Segmentation in Digital Subtraction Angiography\\
}

\author{\IEEEauthorblockN{1\textsuperscript{st} Lemeng Wang}
\IEEEauthorblockA{\textit{School of artificial intelligence} \\
\textit{Beijing University of Posts and Telecommunications}\\
Beijing, China \\
wlm295@bupt.edu.cn}
\and
\IEEEauthorblockN{2\textsuperscript{nd} Wentao Liu}
\IEEEauthorblockA{\textit{School of artificial intelligence} \\
\textit{Beijing University of Posts and Telecommunications}\\
Beijing, China \\
liuwentao@bupt.edu.cn}
\and
\IEEEauthorblockN{3\textsuperscript{rd} Weijin Xu}
\IEEEauthorblockA{\textit{School of artificial intelligence} \\
\textit{Beijing University of Posts and Telecommunications}\\
Beijing, China \\
xwj1994@bupt.edu.cn}
\and
\IEEEauthorblockN{4\textsuperscript{th} Haoyuan Li}
\IEEEauthorblockA{\textit{School of artificial intelligence} \\
\textit{Beijing University of Posts and Telecommunications}\\
Beijing, China \\
email address or ORCID}
\and
\IEEEauthorblockN{5\textsuperscript{th} Huihua Yang}
\IEEEauthorblockA{\textit{School of artificial intelligence} \\
\textit{Beijing University of Posts and Telecommunications}\\
Beijing, China \\
yhh@bupt.edu.cn}
\and
\IEEEauthorblockN{6\textsuperscript{th} Feng Gao}
\IEEEauthorblockA{\textit{Department of Interventional Neuroradiology} \\
\textit{Beijing Tiantan Hospital, Capital Medical University,}\\
Beijing, China \\
13910172189@163.com}
}

\maketitle

\begin{abstract}
Cerebrovascular disease is one of the major diseases facing the world today. Automatic segmentation of intracranial artery (IA) in digital subtraction angiography (DSA) sequences is an important step in the diagnosis of vascular related diseases and in guiding neurointerventional procedures. While, a single image can only show part of the IA within the contrast medium according to the imaging principle of DSA technology. Therefore, 2D DSA segmentation methods are unable to capture the complete IA information and treatment of cerebrovascular diseases. We propose A timing sequence image segmentation network with U-shape, called TSI-Net, which incorporates a bi-directional ConvGRU module (BCM) in the encoder. The network incorporates a bi-directional ConvGRU module (BCM) in the encoder, which can input variable-length DSA sequences, retain past and future information, segment them into 2D images. In addition, we introduce a sensitive detail branch (SDB) at the end for supervising fine vessels. Experimented on the DSA sequence dataset DIAS, the method performs significantly better than state-of-the-art networks in recent years. In particular, it achieves a Sen evaluation metric of 0.797, which is a 3\% improvement compared to other methods.

\end{abstract}


\section{Introduction}
Cerebrovascular disease, with its high incidence, has always been a major health issue worldwide\cite{management}. The IA is complex and fragile, and a sudden break or blockage of blood flow somewhere can lead to insufficient blood supply to the entire brain, causing damage to brain tissue. This condition is known as a stroke and has an extremely high rate of disability and death\cite{strok1}.

The IA comprise one of the cerebrovascular systems within the human body. Alongside the main blood vessels, the intracranial region houses densely branching thin vessels that play a crucial role in transporting blood to different brain tissues. IA are much smaller in volume than the blood vessels in the rest of the body, and when combined with the pressure put on them by the brain tissue, they significantly slow down blood flow\cite{primer}. Cerebrovascular disease usually shows up as abnormalities like vessel narrowing and occlusion, including intracranial arterial stenosis and middle cerebral artery occlusion. It is possible to extract interest regions and ignore interference from other areas by segmenting the structure of the IA, which enables a clear observation of structural abnormalities in the IA. In order to make more accurate diagnoses and prevent them, doctors can use this information to assess the extent of cerebrovascular lesions as accurately as possible\cite{Brain}.

\begin{figure}[t]
\centerline{\includegraphics[scale=0.35]{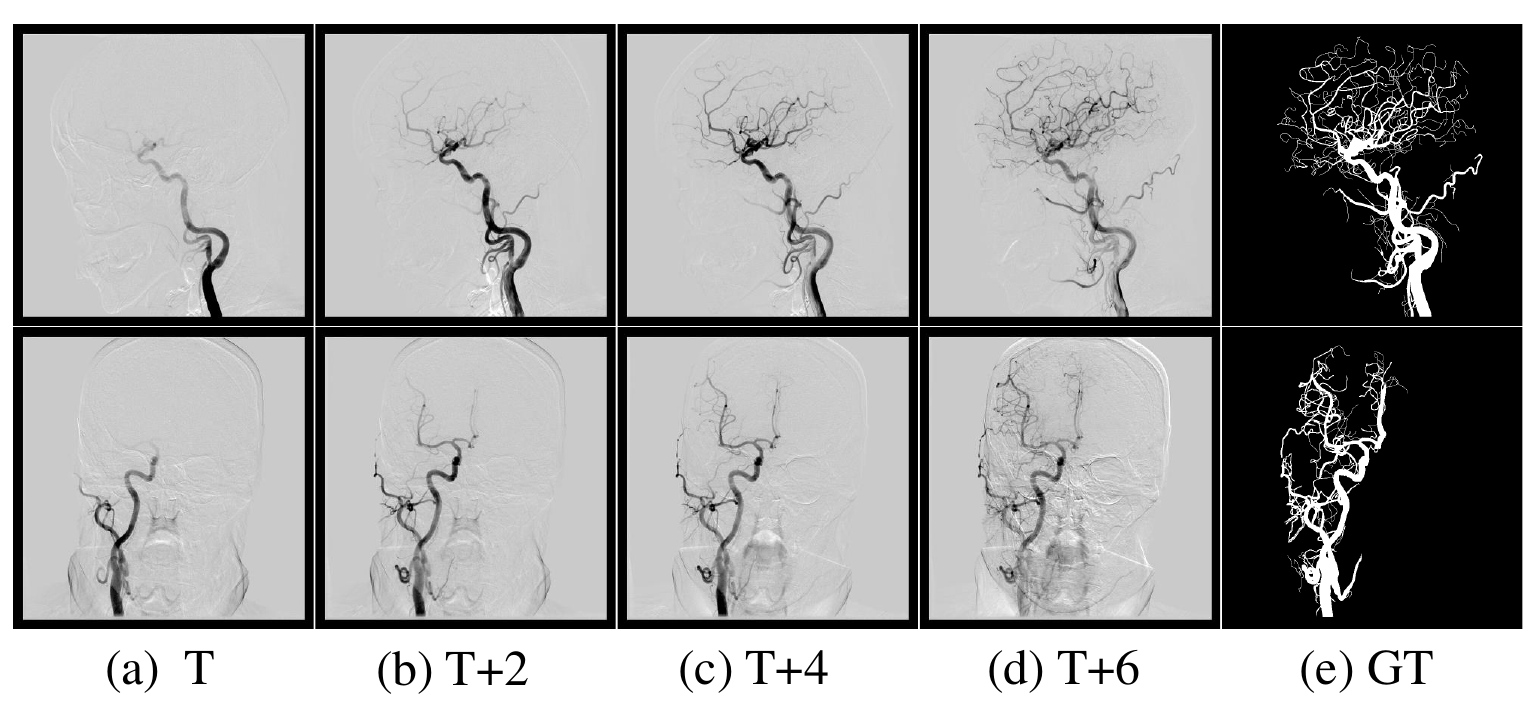}}
\caption{Representative samples of DSA sequences about IA. The first row is a sample of DSA sagittal observations and the second row is a sample of DSA coronal observations. (a)-(d) are arterial continuous frames from T to T + 6 time periods, and (e) is the segmented ground trurh (GT) of the sequence.}
\label{Sequence}
\end{figure}

Early stage of cerebrovascular disease can effectively reduce IA accidents if IA imaging abnormalities are detected in time. Currently, digital subtraction angiography (DSA) remains the gold standard for visualization and diagnosis of angiographic features in most cerebrovascular diseases\cite{DSA}. DSA is a vascular imaging technique. During diagnostic and neurovascular interventions, dynamic imaging of the cerebrovascular by DSA enables observation of blood flow dynamics as well as changes in vessel appearance over time\cite{su}. The recording usually lasts from 3 to 15 seconds, with a sampling rate of 3 to 5 frames per second\cite{dias}. To date, DSA images have mostly been visually examined by neuroradiologists, which can be cumbersome, subjective, and prone to error. Recent studies on segmentation based on DSA images largely adhere to the traditional 2D image segmentation methods, that is, processing a single image extracted from the DSA sequence\cite{DSA2D1,DSA2D2}. Intraoperative DSA examinations for IA-related diseases typically involve sequential images over time, with the position of the contrast agent in the vessel changing at different times, resulting in temporal differences in image visibility. As shown in Fig.~\ref{Sequence}, only a portion of the IA that filled with contrast agents can be displayed in a single image. In fact, successful detection of IA abnormalities depends largely on clear and accurate vascular imaging information in the DSA image sequence. 2D DSA segmentation methods alone are inadequate for the diagnosis and treatment of cerebrovascular diseases due to their failure to capture the complete information related to the IA. The segmentation of DSA sequences exhibiting 2D+time properties should be pursued. Therefore, we consider the segmentation of DSA sequences in IA as a task of dimensionality reduction, where the source consists of 2D+ time images and the target is a 2D mask.

To solve the issue, we propose a \textbf{T}iming \textbf{S}equence \textbf{I}mage segmentation network (TSI-Net) that takes variable frames DSA sequences as input. In TSI-Net, the bi-directional ConvGRU\cite{convgru} module (\textbf{BCM}) are used to model temporal blood flow features, which learn the spatio-temporal correlation of the target from the features of the current image and its neighboring frames. According to the special output mode of BCM, the final output of the network is a 2D image, which realizes the dimensionality reduction segmentation. To improve the sensitivity of the segmentation, a sensitive detail branch (SDB) is employed for precision segmentation of fine vessels. In summary, the contributions in the paper can be summarized as follows:

\begin{itemize}
\item We designed a timing sequence image segmentation network (TSI-Net) by a U-shape. There is a BCM for processing continuous time image information. In BCM, we concatenate multiple Gate Recurrent Unit cells in both forward and reverse order directions, and finally connect them head to tail to form a GRU layer. The update and reset gates of each ConvGRU cell can learn and forget the contents of the current image memory in the previous or next frame, respectively.
\item At the output side we employ an SDB to perform auxiliary supervision of the network, which is used to improve the sensitivity and connectivity of the segmentation.
\item The experimental results demonstrate the necessity of BCM and the effectiveness of SDB. Comparative results on DIAS with other state-of-the-art methods validate the promising segmentation performance of the proposed approach.
\end{itemize}

\begin{figure*}[t]
\centerline{\includegraphics[scale=0.4]{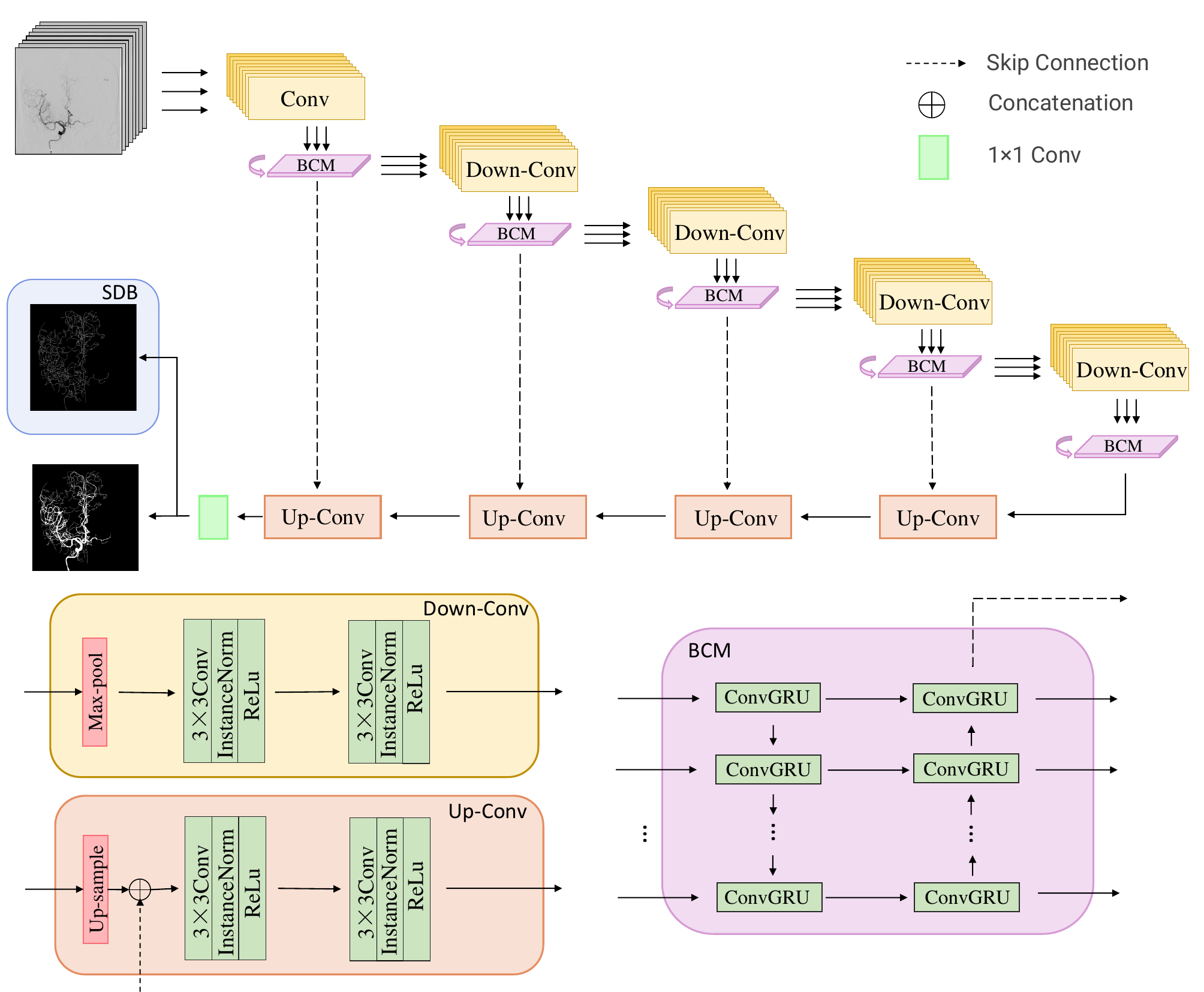}}
\caption{Network architecture of the TSI-Net for IA segmentation in DSA. The inputs are 2D+time DSA sequences of different lengths. The output is a 2D segmented image.}
\label{TSI-Net}
\end{figure*}

\section{Related Work}
Thanks to the development of biomedical technology, we can easily access a large amount of biological and physiological data through modern equipment. At the same time, the rapid development of deep learning has brought new ideas to solve the problem of IA segmentation, assisting doctors to realize the extraction and quantification of vascular indicators for clinical diagnosis.

The groundbreaking U-Net developed by Ronneberger~\emph{et al.}\cite{UNet} is considered a pioneering method in the field of segmentation and has become the focal point of medical image segmentation due to its outstanding performance. As a convolutional neural network, derivatives of U-Net have been widely applied to various vascular segmentation tasks. For example, 2D\cite{resdo,mcunet,meng} or 3D\cite{ERNet,spider} methods with modifications made mainly to the U-Net backbone network to achieve better segmentation results. They use the structure of an encoder-decoder and introduce popular structural components such as attention mechanism\cite{ERNet,resdo,multiscale}, atrous convolution\cite{mcunet,amsunet}, gate structure\cite{multiscale,spider}. Liu~\emph{et al.}\cite{MIP} combined the original 3D U-Net with Maximum Intensity Projection and used a spatial attention-guided 3D Inception U-Net segmentation flow and a 2D composite multi-directional MIPs U-Net segmentation flow to segment 3D MRA IA images. Zhang~\emph{et al.}\cite{mcunet} designed a cascade of modules by combining the spatial attention module with the dense atrous convolution module and multi-kernel pooling module to obtain a larger receptive field for retinal vessel segmentation. Liu~\emph{et al.}\cite{multiscale} used convolutional long short-term memory network and deep dilated convolution in the decoder for retinal vessel segmentation. Lee~\emph{et al.}\cite{spider} considered the connectivity of vessels between axial slices to achieve MR modality IA segmentation and inserted long short-term memory (LSTM) into the model to capture the context of continuous data.

However, there is limited research on DSA vascular segmentation. Meng~\emph{et al.}\cite{meng} proposed a CNN-based segmentation framework that includes a multi-scale atlas convolution module, improved dense blocks, and redesigned skip connections for the segmentation of IA in DSA images. Zhang ~\emph{et al.}\cite{BFFD} proposed a boundary enhancement and feature denoising module to improve the network's ability to extract boundary information in coronary artery DSA segmentation. These two methods were developed on an internal DSA dataset and are only applicable to 2D DSA vascular segmentation. However, DSA is a modality with powerful spatio-temporal visualization of blood flow dynamics in continuous frames, and these methods fail to consider the temporal characteristics of DSA imaging patterns and do not fully utilize the temporal and spatial correlation in IA DSA images.

\section{Method}
The automatic IA segmentation of DSA sequences is an unprecedented and challenging task. We propose a timing sequence image segmentation network called TSI-Net, and its structure is shown in Fig.~\ref{TSI-Net} 

\subsection{TSI-Net}
The TSI-Net architecture consists of an encoder-decoder network based on the U-shape design. The encoder of TSI-Net receives preprocessed IA DSA sequences as input. In contrast to the original U-Net architecture, the DSA sequences are fed to the network frame by frame. The bi-directional ConvGRU module (\textbf{BCM}) is utilized to incorporate temporal information features, where each group of Down-Conv shares one weights. The encoder consists of four IA down-sampling layers, and each frame of one sequence undergoes two convolution in the same layer and then connect into BCM block frame by frame. Following the bi-directional ConvGRU module (\textbf{BCM}), the output sequence undergoes IA down-sampling through maximum pooling with a step size of 2, and the output of the final ConvGRU cell is connected to the decoder using a skip connection. 

Each IA down-sampling doubles the number of channels in the feature map, which increases the receptive field of the deep features and ensures their global nature to some extent. The decoder is symmetric to the encoder and consists of 4 up-sampling layers. Connected the up-sampling feature maps with the same resolution feature maps from encoder can reduce the loss of edge features because of up-sampling. The features of the encoder have higher resolution, while they extracted by down-sampling, this information is not retrievable at the time of up-sampling. Finally, a 1x1 fully connected layer is used to classify the background and blood vessels in the image to output a segmented image of IA. The number of channels start from 32. An important improvement over U-Net is the use of 2D+T DSA sequences that contain temporal information instead of a single 2D image as an input.

\subsection{Bidirectional ConvGRU Module}
It is essential to depict the sequential evolution of blood flow contrast information in the temporal domain, Since a DSA sequence presents a series of IA information in chronological order. Our proposal involves utilizing the BCM to simulate the temporal progression of blood flow appearance in the sequence. ConvGRU\cite{convgru} is an extension of the traditional fully connected GRU, with convolutional structures in both the input-to-state and state-to-state connections. Define the given sequence with t-frames as S = ($f_1$,$f_2$,...,$f_t$), and $h_t$ represents the hidden state at time t. A ConvGRU cell consists of a reset gate and a update gate are expressed as $R_t$ and $U_t$. Through these gates, ConvGRU cells can not only achieve selective memory but also forgetting. Based on the above definition, the overall update process at the ConvGRU cell for different time steps in a DSA sequence is as follows:

\begin{equation}
\begin{split}
\mathcal{U}_{t}=\sigma({\omega}_{u}*{f}_{t}+{\omega'}_{u}*{h}_{t-1})\\
\mathcal{R}_{t}=\sigma({\omega}_{r}*{f}_{t}+{\omega'}_{r}*{h}_{t-1})\\
{\tilde{h}}_{t}=\tanh({\omega}*{f}_{t}+\mathcal{R}_{t}\circ({\omega'}*{h}_{t-1}))\\
{h}_{t}=(1-\mathcal{U}_{t})\circ{\tilde{h}}_{t}+\mathcal{U}_{t}\circ{h}_{t-1}
\end{split}
\end{equation}

where `*' denotes the convolutional operation, `$\circ$' represents the hadamard product, $\sigma(.)$ denotes sigmoid function, $\omega$, $\omega'$ means weights that can be learned. $\tilde{h}$ is a candidate activation which is computed in a manner similart to that of a traditional recurrent unit in an Recurrent Neural Network. The formula simplifies the expression and ignores the bias term. 

We input the DSA sequences frame-by-frame into encoder for feature extraction. Inspired by\cite{bigru}, we stack two sets of ConvGRU to compose BCM in forward and backward propagation, respectively, to enhance the exchange of spatio-temporal information between them. In this way, the BGM not only remembers past sequences but also predicts future sequences. This can be expressed as:

\begin{equation}
\begin{split}
{h}_{t}^{f}=ConvGRU({h}_{t-1}^{f},{f}_{t})\\
{h}_{t}^{b}=ConvGRU({h}_{t+1}^{b},{h}_{t}^{f})\\
{h}_{t}=\tanh({\omega}_{f}*{h}_{t}^{f}+{\omega}_{b}*{h}_{t}^{b})
\end{split}
\end{equation}

where ${h}_{t}^{f}$ and ${h}_{t}^{b}$ represent the hidden state from forward and backward ConvGRU cells, respectively. $l{h}_{t}$ represents the final output of BCM.

It should be noted that all blocks in the BCM share the same weights. The last layer of each BCM aggregates the features of all frames in the DSA sequence. This operation is designed to allow the network to learn the relationships between frames rather than treating them as individual frames for processing.

\subsection{Sensitive Detail Branch}
Data imbalance is a common problem in medical image segmentation. The region of inerest in a medical image is usually much less than entire image. Moreover the proportion of voxels in the regions to be segmented is smaller than the rest of the image. If the data is unbalanced, the learning process may converge to a suboptimal local minimum, and the final prediction results could be biased towards non-relevant features. This data imbalance issue is particularly severe in vessel segmentation. 

Consequently, we used sensitive detail branch (SDB) for adjunctive monitoring of the thin vessels. In this branch, we use the skeletonization operation to perform centerline extraction of the original ground truth (GT) as a new label. In this module for assisted supervision, we calculate the mean square error values between the actual and predicted values one by one. The aim is to keep reducing the gap between the two, making the network more sensitive to fine vessels and some details, and maintaining a higher connectivity between the vessels.It can be defined as:

\begin{equation}
\mathcal{L}_{SDB}=(pre_i - \phi(gt_i))^2
\end{equation}

where $pre_i\in[0,1]$ represent predicted probability of network output, $gt_i\in[0,1]$ denotes GT, $\phi(gt_i)$ is the centerline lifting method using the skeletonization operation. 

On the one hand, the most often used loss function is the cross-entropy loss, which examines each pixel individually and compares the class prediction vectors to the GT. It can be defined as:

\begin{equation}
\mathcal{L}_{ce}(pre_i,{gt}_i)=\sum_{i=1}^{N}[(pre_i-1)\times log(1-gt_i)-pre_i\times log(gt_i)]
\end{equation}

where N is the total pixel number.

On the other hand, the dice coefficient is a measure of the overlap area of the picture difference region and is used to compare the predicted map to the GT. It is a superior metric for minor goals. The dice loss ($\mathcal{L}_{dice}$) can be defined as:

\begin{equation}
\mathcal{L}_{dice}(pre_i,{gt}_i)=1-\frac{2\sum_{i=1}^{N} pre_i\times gt_i}{\sum_{i=1}^{N}pre_i+\sum_{i=1}^{N}gt_i}
\end{equation}

The final loss function is defined as:

\begin{equation}
\mathcal{L}_{total}=\mathcal{L}_{ce}+\lambda_1\mathcal{L}_{dice}+\lambda_2 \mathcal{L}_{SDB}
\end{equation}

where $\lambda_1$ and $\lambda_2$ denotes the weight of $\mathcal{L}_{dice}$ and $\mathcal{L}_{SDB}$ respectively.

Auxiliary supervised mean square error loss improves the sensitivity of the model. Sensitivity measures how many true positives the model correctly detects among all actual positives. As a result, it allows for better segmentation of the end vessels and fine vessels of the vascular tree, improving segmentation accuracy and vascular connectivity.

\begin{figure*}[t]
\centerline{\includegraphics[scale=0.5]{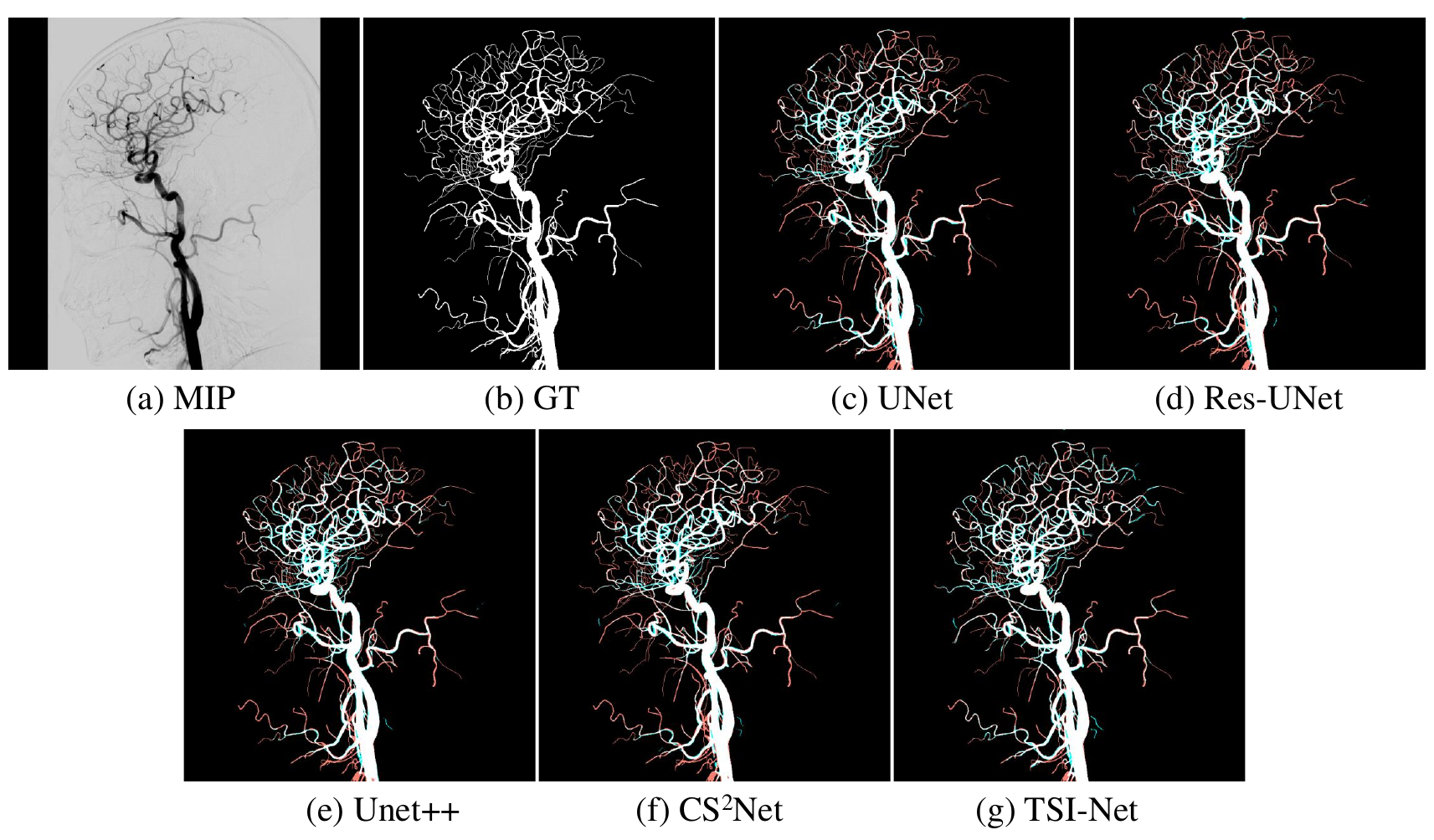}}
\caption{Visualization of IA segmentation results. (a) is the MIP plot of the IA DSA sequence in the DIAS dataset (b) is the GT of (a). (c) is the segmentation result of UNet, (d) is the segmentation result of Res-UNet, (e) is the segmentation result of UNet++, (f) is the segmentation result of CS$^2$Net, and (g) is the segmentation result of our network. The red pixel indicates false positives, and the green pixel indicates false negatives.}
\label{result}
\end{figure*}

\section{Experiment}
\subsection{Data and Evaluation}
To validate the performance of our proposed TSI-Net on the IA segmentation task, we conducted experiments on the DIAS dataset\cite{dias}.

The DIAS dataset contains 60 one-on-one pixel-level labeled sequences. In the same way as the data are divided in \cite{dias}, 30 as training samples, 10 as validation samples, and 20 as test samples. The DSA sequences have dimensions of $800\times800\times8$. While $800\times800$ is the resolution of a single frame of the image in the sequence, and 8 is the number of frames in a single sequence.

We compare the predicted segmentation results with the corresponding ground truth to calculate the dice similarity coefficient (Dice) and evaluate the accuracy (Acc), sensitivity (Sen), specificity (Spe),  and intersection over union (IOU) of the binary segmentation maps obtained using thresholding. Their definitions are as follows:

\begin{equation}
Dice=\frac{2TP}{2TP+FP+FN} 
\end{equation}
\begin{equation}
Acc=\frac{TP+TN}{TP+TN+FP+FN} 
\end{equation}
\begin{equation}
Sen=\frac{TP}{TP+FN} 
\end{equation}
\begin{equation}
Spe=\frac{TN}{TN+FP} 
\end{equation}
\begin{equation}
IOU=\frac{TP}{TP+FP+FN} 
\end{equation}

where true positives (TP) and true negatives (TN) denote the numbers of correctly segmented vascular pixels and nonvascular pixels, respectively; false positives (FP) and false negatives (FN) denote the numbers of incorrectly segmented vascular pixels and nonvascular pixels, respectively.

Additionally, we introduced the vascular connectivity (VC) metric\cite{dias}. This metric calculates the ratio of connected components in the segmentation image to those in the ground truth, providing a measure of vessel connectivity.

\subsection{Implementation Details}

Firstly, all DSA sequences were z-score normalized. We randomly crop the input sequences into patches of size $8\times64\times64$. We randomly crop the input image into patches of size $8\times64\times64$. Randomized horizontal, vertical and 90-degree flips were used for data enhancement. The preprocessed sequences are used for training with an epoch size of 64. We optimize the model parameters using AdamW, with an initial learning rate of 5e-4. The learning rate is gradually reduced using the cosine annealing algorithm for more than 100 iterations. All experiments are conducted using PyTorch with constant hyperparameters and on a single GeForce RTX 3090 GPU.

\section{Result}
\subsection{Comparisons With the State-of-The-Art Methods}

We conducted IA segmentation experiments on the most popular networks, including U-Net\cite{UNet}, Res-UNet\cite{ResUNet}, UNet++\cite{UNet++} and CS$^2$Net\cite{cs2-net}. To address the limitation of network processing only individual 2D images, a dimension reduction module (DRM)\cite{dias} is incorporated to handle the DSA sequence. Specifically, given a DSA sequence $s\in\mathbb{R}^{C\times F\times H\times W}$, where $C$, $F$, $H$ and $W$ denote the number of channels, frame, height, and width, respectively. After DRM compression, the initial channel number is changed to 1. The compressed input is then fed into the 2D model, where property F replaces C as the input channel:
\begin{equation}
Out_{DRM} = \mathcal{M}(Squ(s))
\end{equation}
\begin{table}[h]
\centering
\caption{Comparison of TSI-UNet with the State-Of-The-Art methods on stare} \label{table:result}
\renewcommand\arraystretch{1.5}
\setlength{\tabcolsep}{1.3mm}{
\begin{tabular}{l|ccccccc}
\hline
\multirow{2}{*}{Methods} & \multicolumn{7}{c}{Metrics}                                                                                                  \\ \cline{2-8} 
                         & Dice            & AUC             & Acc             & Sen             & Spe             & IOU             & VC↓              \\ \hline
UNet                     & 0.7633          & 0.9797          & 0.9637          & 0.7601          & 0.9818          & 0.6198          & 79.60          \\
Res-UNet                 & 0.765           & 0.9787          & 0.9644          & 0.7593          & 0.9826          & 0.6217          & 73.05          \\
UNet++                   & 0.7643          & 0.9808          & 0.9645          & 0.7537          & \textbf{0.9832} & 0.6211          & 68.68           \\
CS2Net                   & 0.769           & \textbf{0.9817} & 0.9648          & 0.7663          & 0.9826          & 0.6272          & 68.40           \\
TSI-Net                  & \textbf{0.7861} & 0.9816          & \textbf{0.9664} & \textbf{0.7971} & 0.9815          & \textbf{0.6492} & \textbf{42.70} \\ \hline
\end{tabular}}
\end{table}

where, $Squ$ is dimensionality squeeze operation and $\mathcal{M}$ is a 2D model. 

Fig.~\ref{result} shows the segmentation results. We perform minimizing operations on the entire sequence for each pixel along the sequence dimension to obtain minimum intensity projection (MIP) images. It can be seen that all models have comparable segmentation ability for large blood vessels, while for small blood vessels, there is a significant reduction of red pixels in the segmentation results of TSI-Net, which indicates that our network is more powerful for fine vessel segmentation.

The qualitative segmentation results for the DIAS dataset are presented in Table~\ref{table:result}. Overall, TSI-Net demonstrates the best performance on this dataset, exhibiting an approximately 2\% improvement in Dice coefficient, which is outperforming to state-of-the-art methods.

Notably, TSI-Net exhibited a noticeable 3\% improvement in the Sen score and a significant reduction in the VC compared to other methods, while only experiencing a marginal 0.1\% decrease in Auc. These enhancements imply an improved ability of the model to accurately segment fine vessels and ensure improved vessel connectivity with fewer disruptions. Precisely capturing vessel distribution in interventional procedures serves as a crucial benchmark, and both metrics hold particular significance in scenarios characterized by highly imbalanced background and target distributions.

However, in contrast to the other composite evaluation metrics, the Spe is relatively low. Sen and Spe represent the proportions of correctly predicted positive and negative pixels, respectively. During the extraction of weakly vascularized pixels, TSI-Net introduces some false positives, indicating incorrect predictions of non-vascular pixels. We believe that the model further improves its ability to extract weak blood vessels while naturally increasing the acceptance of false positives.

\begin{table*}[t]
\centering
\caption{Ablation study of TSI-Net with different network configurations in IA segmentation (baseline: UNet , UCM :uni-directional ConvGRU module, SDB :sensitive , detail branch)} \label{table:ablation}
\renewcommand\arraystretch{1.5}
\setlength{\tabcolsep}{2mm}{
\begin{tabular}{cccc|ccccccc}
\hline
\multicolumn{4}{c|}{\textbf{Method}}                           & \multicolumn{7}{c}{\textbf{Metrics}}                                                                                                              \\ \hline
\textbf{Baseline} & \textbf{UCM} & \textbf{BCM} & \textbf{SDB} & \textbf{Dice}   & \textbf{AUC}                           & \textbf{Acc}    & \textbf{Sen}    & \textbf{Spe}    & \textbf{IOU}    & \textbf{VC$\downarrow$}    \\ \hline
\checkmark                 &              &              &              & 0.7633          & 0.9797                                 & 0.9637          & 0.7601          & 0.9818          & 0.6198          & 79.60          \\
\checkmark                 & \checkmark            &              &              & 0.7851          & 0.9849                                 & \textbf{0.9681} & 0.7549          & \textbf{0.9868} & 0.6478          & 64.32          \\
\checkmark                 &              & \checkmark            &              & 0.7842          & {\color[HTML]{0D0D0D} \textbf{0.9854}} & 0.9671          & 0.7729          & 0.9844          & 0.6468          & 54.49          \\
\checkmark                 &             & \checkmark            & \checkmark            & \textbf{0.7861} & 0.9816                                 & 0.9664          & \textbf{0.7971} & 0.9815          & \textbf{0.6492} & \textbf{42.70} \\ \hline
\end{tabular}}
\end{table*}

\begin{figure}[h]
\centerline{\includegraphics[scale=0.3]{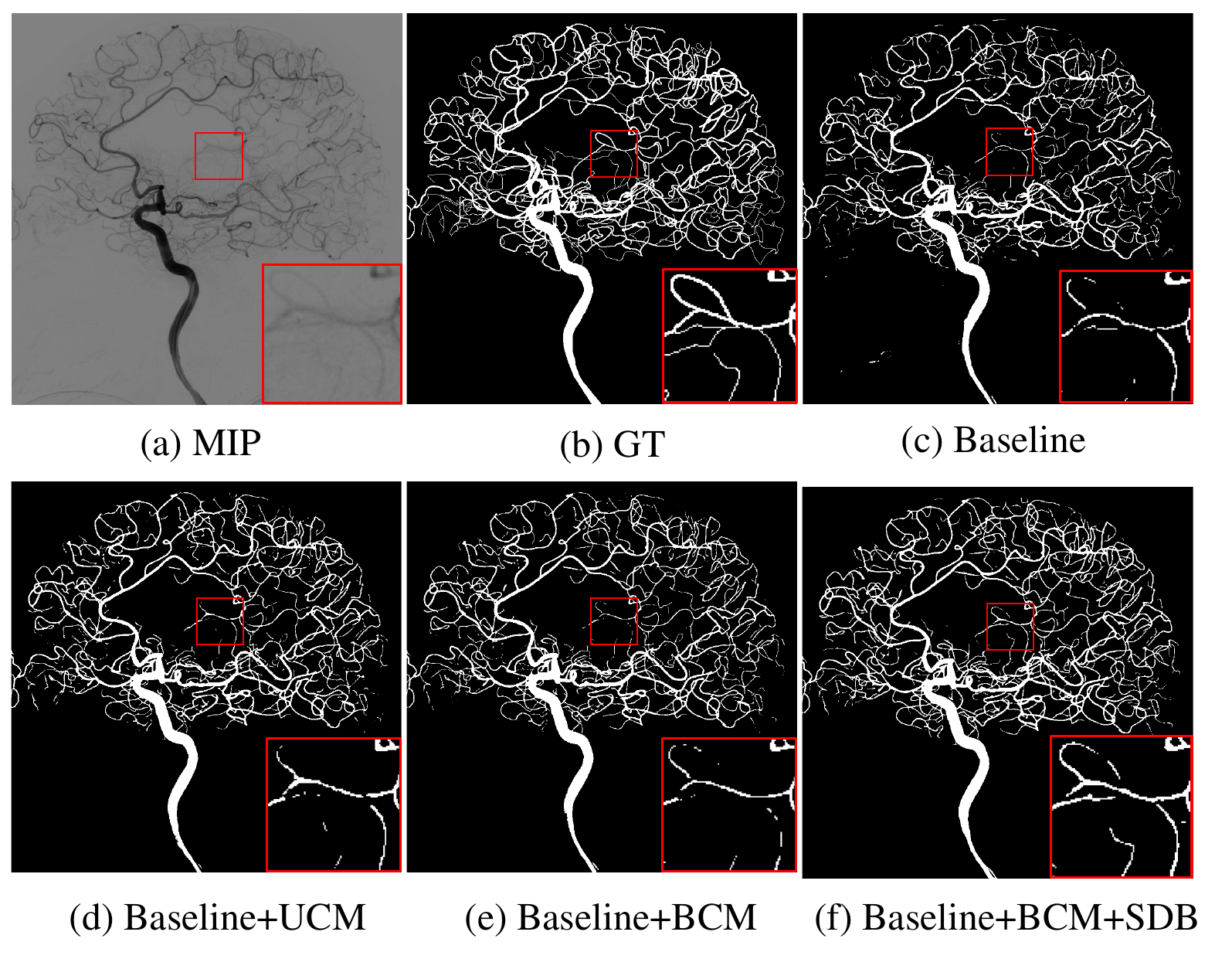}}
\caption{Visual comparison of ablation studies of key components in TSI-Net. (a) is the MIP plot of the IA DSA sequence in the DIAS dataset. (b) is the GT of (a). (c)-(f) are segmentation results for baseline, baseline+UCM, baseline+BCM, and baseline+BCM+SDB (ours), respectively. Ucm denotes uni-directional ConvGRU module, BCM denotes bi-directional ConvGRU module, and SDB is the sensitive detail branch. To show the details, we enlarge the red highlighted rectangular regions in the images.}
\label{ablation}
\end{figure}

\subsection{Ablation Studies}
To validate the effectiveness of the components, we conducted ablation experiments and evaluated the impact of each component on the results on the DIAS dataset. Using UNet as the baseline for the experiments, We incrementally incorporated Uni-directional ConvGRU Module (UCM) ,BCM and SDM to the baseline of the ablation study. All experiments were conducted using the same hyperparameter configuration. In our ablation study, all competitors were run on the same computing environment with the same data augmentation to ensure a fair comparison. Again since UNet cannot handle DSA sequences, the initial number of channels is set to 1 and compressed by DRM. The compressed input is then fed into the 2D model where "F" replaces "C" as the input channel. The results are shown in the Table~\ref{table:ablation}. 

The inclusion of ConvGRU resulted in significant improvements in all model scores compared to the baseline, with bi-directional connectivity demonstrating superior performance over uni-directional connectivity across multiple metrics. When compared to baseline + BCM, the addition of SDB further enhanced the Sen score by 2.4\%, achieving a maximum score of 0.7971. Auc and Spe exhibited only a marginal decrease of approximately 0.4\%. Considering the degree of imbalance between IA vascular and background and practical medical application scenarios, the noted improvement in Sen and VC indicates that the model facilitates enhanced preservation of fine vascular information and vascular topology, ultimately enhancing the performance of sequence segmentation.

The ablation results were further visualized, as depicted in Fig~\ref{ablation}, which includes an IA sequence featuring numerous fine vessels. Consequently, we zoomed in on specific image details to facilitate clearer visualization, focusing on the highlighted rectangular area. Within the zoomed subfigure, the thin vessels in the original image exhibit low contrast, appearing blurred and challenging to discern by the human eye. The expert has labeled the blood vessels in the GT, and based on the visualization of the segmentation results, it can be observed that Baseline, Baseline+UCM, and Baseline+BCM show incomplete extraction of the thin blood vessels in this area. However, the inclusion of Baseline+BCM+SDB resolves this issue by introducing fewer false-negative pixels, thereby visibly enhancing vascular connectivity.

\section{Conclusion}
We propose a spatio-temporal segmentation method for IA based on a timing sequence image segmentation network. In this network, the incorporation of BCM enables the network to consider both past and future information to better learn IA sequences that contain temporal information. Unlike previous segmentation networks, this network takes 2D+time DSA sequences as input and outputs 2D segmented images. Evaluation metrics show that TSI-Net achieves excellent overall performance compared to other state-of-the-art methods. In addition, the visualization of IA segmentation results with the addition of SDB demonstrated the power of TSI-Net in extracting fine vessels and improving vascular connectivity. In the future, we will continue to explore this work to improve network sensitivity while being able to achieve better overall correctness.

\bibliographystyle{IEEEtran}
\bibliography{refer}

\end{document}